# Binary GH Sequences for Multiparty Communication

Krishnamurthy Kirthi

**Abstract**
This paper investigates cross correlation properties of sequences derived from GH sequences modulo p, where p is a prime number and presents comparison with cross correlation properties of pseudo noise sequences. For GH sequences modulo prime, a binary random sequence B(n) is constructed, based on whether the period is p-1 (or a divisor) or 2p+2 (or a divisor). We show that B(n) sequences have much less peak cross correlation compared to PN sequence fragments obtained from the same generator. Potential applications of these sequences to cryptography are sketched.

**Introduction**
Spread spectrum systems provide secure communications by spreading a signal over large frequency band. They are implemented using short periodic pseudo-random sequences with good correlation properties [1]-[3]. The sequences necessary for direct-sequence spreading and spread spectrum analysis must have low cross correlation characteristics. Peak cross-correlation is used when measuring the difference between two sequences of different time series [4].

In a recent paper [5], GH sequences that are related to Fibonacci sequences [6] were shown to have good pseudo-randomness properties. As is well-known randomness can be examined both from a physics perspective [7]-[10] as well as an algorithmic one [11]-[14]. The family of GH sequences presents a new take on an old mathematical structure and is therefore of much interest. These sequences can be used also in sending side information that can help in authentication in cryptography systems which is one of the central problems in a networked society [19]-[27]. In particular, good random sequences can be of help in frustrating man-in-the-middle attacks in P2P systems.

In this paper, we present the cross correlation of binary GH sequences and compare these to that of PN sequences. For this comparison, we use peak cross correlation function as a measure and show that GH sequences score over PN sequences. These sequences can have applications in cryptography in authentication of parties, especially in P2P systems.

**Periods of shift register and Fibonacci sequences**

A PN sequence is a periodic binary sequence generated using linear feedback shift register (LFSR) structure. Figure 1 shows a fragment of a PN sequence generated using LFSR structure for the polynomial $p(z)=z^{45}+z^4+z^3+z+1$. Here we show 63 bits of a sequence whose period is $3.52 \times 10^{13}$.

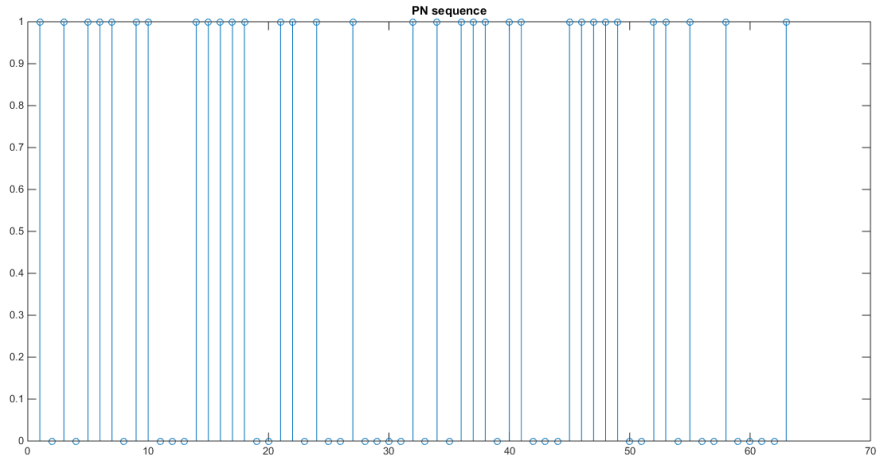

Figure 1: PN sequence fragment of length 63

The period of a PN sequence produced by a linear feedback shift register with m stages cannot exceed $2^m-1$. When the period is exactly $2^m-1$, the PN sequence is maximal length sequence or m-sequence.

The periods of the GH sequences modulo *m* will be identical to that of the corresponding Fibonacci sequence [15]. Consider a GH sequence to mod *p* where *p* is a prime number. It is quite clear that there can at most be $p^2-1$ pairs of consecutive residues in a period. Since $p^2-1$ is (p-1)(p+1), the period of the residue sequence will either be divisor of p-1 or p+1 (or equivalently of 2p+2) [5].

The periods of GH sequences mod *p*, with *p* as a prime, has been shown to be either p-1 if p≡1 or 9 (mod 10) or 2p+2 if p≡3 or 7(mod 10). Specifically, we have:

(i) (p-1) or a divisor thereof if the prime number *p* ends with 1 or 9.
(ii) (2p+2) or a divisor thereof if the prime number *p* ends with 3 or 7.
(iii) 20 for *p* =5.

The pertinent result for the period, N, of GH sequence for non-prime modulo *m* is:

$N(m) \leq 6m$ with equality iff $m = 2 \times 5^n$. Thus, the periods can be grouped into either p-1 or 2p+2. This is the most significant fact from the perspective of generation of random sequences and in [5] it was used to generate a binary random sequence which was shown to have good autocorrelation properties. We want to go beyond that result and determine the cross correlation properties of fragments of this sequence obtained from different regions.

**Cross Correlation properties**

The cross correlation between two sequences is the complex inner product of the first sequence with a shifted version of the second sequence which indicates if the two sequences are distinct. The correlation properties of the sequences are used to detect and synchronize the communication process.

We assign periods of GH sequence mod $p$, with multiples of (p-1) or divisors, as binary value +1 and periods with multiples of (2p+2) or divisor, as binary value -1 and thus call the resulting binary sequence as B(n). The first 20 bits of B(n) are -1,1,-1,1,-1,-1,1,-1,1,1,-1,1,-1,-1,-1,1,1,-1,1 and -1 .

Let us consider prime moduli and determine periodic cross correlation properties of B(n) to determine how good they are from the point of view of randomness. The cross correlation function is calculated using the formula:

$$CCF(k) = \frac{1}{N}\sum_{j=0}^{N-1} A_j B_{j+k}$$

where $A_j$ and $B_{j+k}$ are the binary values of two sequences at different time intervals and N is the length of sequence or period of sequence. The peak cross correlation function value of a cross correlated sequence will be denoted by $CCF_{peak}$.

Figures 2 and 3 present the normalized cross correlation function of the B(n) sequence for 100 and 200 bits.

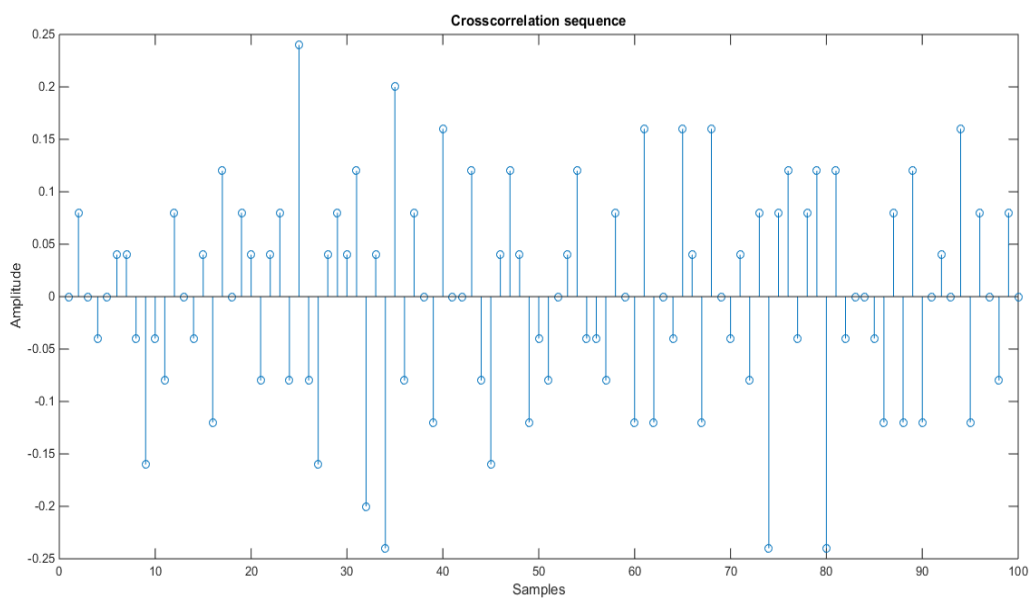

Figure 2. CCF of Binary GH sequences for 100 bits

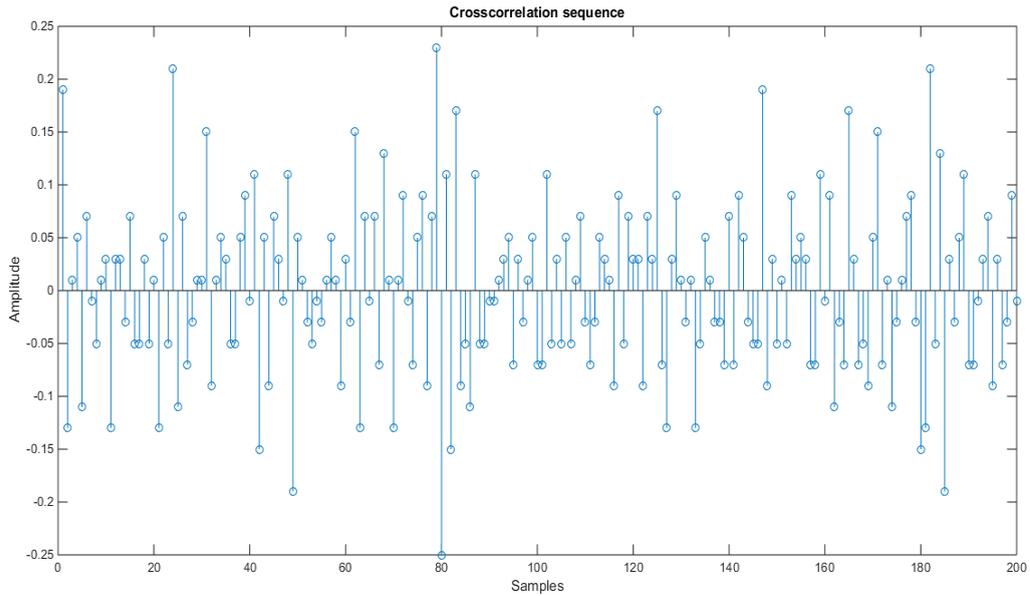

Figure 3. CCF of Binary GH sequences for 200 bits

Looking at Figure 2 and Figure 3, their peak cross correlation values are noted to be 0.25 in both cases.

Figures 4 and 5 present the normalized cross correlation function characteristics of the PN sequences for 100 and 200 points. The PN sequence fragments in these figures are from the expansion of the polynomial $p(z)=z^{45}+z^{4}+z^{3}+z+1$

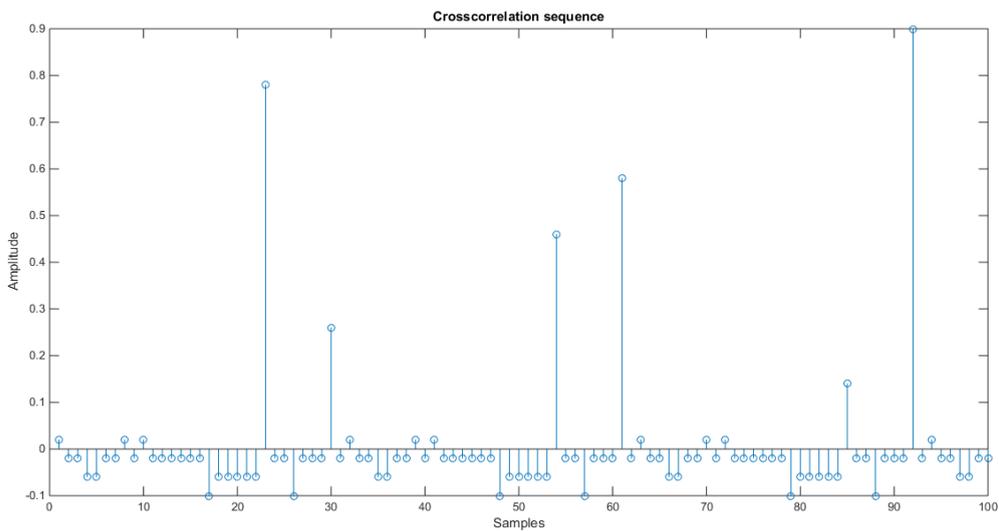

Figure 4. CCF of PN sequences for sequence length 100

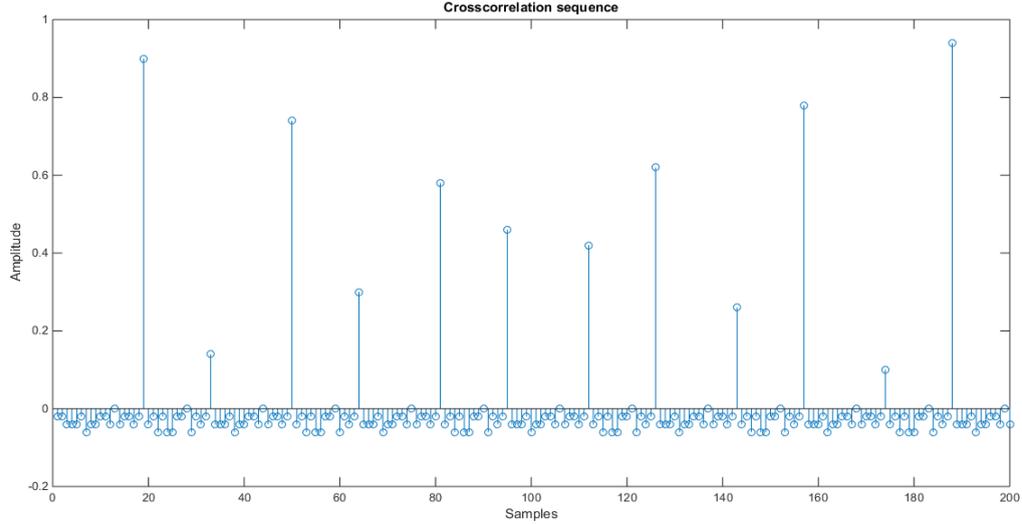

Figure 5. CCF of PN sequences for sequence length 200

Randomness may be calculated using the randomness measure, R(x), of a discrete sequence x by the expression below:

$$R(x) = 1 - \frac{\sum_{k=1}^{n-1}|CCF(k)|}{N-1}$$

where CCF(k) is the cross correlation function value for k and N is the period of sequence to characterize the randomness of a sequence. According to above formula, the randomness measure of 1 indicates that the sequence is fully random whereas randomness measure of 0 indicates a constant sequence. The randomness measure for Figure 2 and Figure 3 are found to be 0.9212 and 0.9342 respectively.

**Comparison between Binary GH sequences and PN sequences**
The following table gives a comparison between peak cross correlation function (CCF) value of Binary GH Sequences and PN Sequences. Here we consider two types of PN sequences: (i) where the fragments are taken from the same long PN sequence; (ii) where the fragments come from different PN sequences.

Table 1: Peak CCF

| Length of bits | Peak CCF value for Binary GH Sequences | Peak CCF value for PN Sequences of same generator polynomials | Peak CCF value for PN Sequences of different generator polynomials |
|---|---|---|---|
| 25 | 0.52 | 0.6 | 0.36 |
| 50 | 0.32 | 0.72 | 0.32 |
| 100 | 0.25 | 0.9 | 0.36 |
| 150 | 0.24 | 0.9 | 0.32 |
| 200 | 0.24 | 0.94 | 0.25 |

As is to be expected the peak cross correlation function of the PN sequences obtained from different generators is lower than that obtained from the same generator. The graph that represents the above table is shown below.

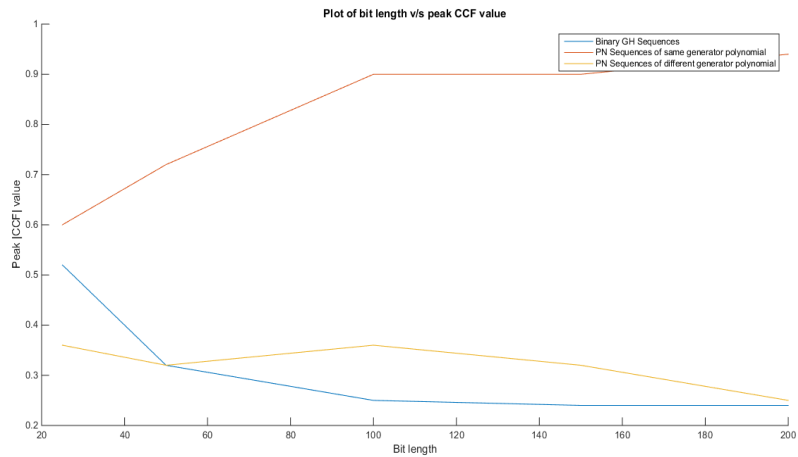

Figure 6. CCF of GH and PN sequences for lengths up to 200

Hence, comparing the peak cross correlation function (CCF) value of GH series mod p with that of pseudo-noise sequences, it is found that GH series mod p has excellent cross correlation properties that is comparable to PN sequences obtained from different generators.

**Conclusion**

We have compared the cross correlation of GH sequences with that of PN sequences and shown that the GH sequences for prime modulo have much better cross correlation properties compared to that of pseudo noise sequences obtained from the same generator. Hence, they are good candidates for cryptography since the generation process is not computationally complex.

GH sequences can be used to convey side-information to help communicating parties to authenticate each other. Their use can therefore be made a part of handshake protocols.